\documentclass[aps,prd,preprint,nofootinbib]{revtex4}
\usepackage{amsmath,amssymb,graphics,epsfig,multirow}%,mathbbol,graphicx,psfrag}
\usepackage{verbatim}

\newcommand{\newc}{\newcommand}
\newc{\bsym}{\boldsymbol}
\newc{\mrm}{\mathrm}
\newc{\ovl}{\overline}
\newc{\ovla}{\overleftarrow}
\newc{\ovra}{\overrightarrow}
\newc{\ra}{\rightarrow}
\newc{\lra}{\leftrightarrow}
\newc{\wtil}{\widetilde}
\newc{\eps}{\epsilon}
\newc{\hc}{\dagger}
\newc{\pd}{\partial}
\newc{\SL}{\!\!\!/}
\newc{\LH}{\hat{L}}
\newc{\RH}{\hat{R}}
\newc{\sWsq}{\sin^2\theta_\mathrm{W}}
\newc{\cWsq}{\cos^2\theta_\mathrm{W}}
\newc{\half}{\frac{1}{2}}
\newc{\hh}{\hat{H}}
\newc{\hphi}{\hat{\Phi}}
\newc{\nonr}{\nonumber}

\newcommand{\set}[1]{\mathbb{#1}}  %% Math blackboard-style bold for 'special' sets
\newcommand{\ud}{\text{d}}  %% Roman font d for differentials

\DeclareMathOperator{\Tr}{Tr}  %% All operators should be Roman, Trace not predefined
\DeclareMathOperator{\diag}{diag}

\graphicspath{{/}{Figures/}}

\begin{document}

\title{An Unusual Two Higgs Doublet Model from Warped Space}

\author{We-Fu Chang}
\email{wfchang@phys.nthu.edu.tw}
\affiliation{Department of Physics, National Tsing Hua University, HsinChu 300, Taiwan}

\author{John N. Ng}
\email{misery@triumf.ca}
\author{A.P. Spray}
\email{aps37@triumf.ca}
\affiliation{Theory Group, TRIUMF, 4004 Wesbrook Mall, Vancouver, B.C., Canada}

\date{\today}

\begin{abstract}
We study a simple two Higgs doublet model (2HDM) in the Randall-Sundrum scenario,
 with an IR brane localized Higgs field and a second doublet arising from a $t\bar{t}$
 condensate due to strong Kaluza-Klein gluon effects.
 The effective 2HDM predicts that the ratio of the brane to condensate vacuum expectation values $\tan \beta \sim 3$.
 It also predicts a standard model like Higgs boson of mass ${\mathcal{O}} (100)$ GeV and a heavier scalar at the scale of the lowest KK gluon mass, which we take to be $M_{KK} \gtrsim 1.5$~TeV.  The pseudoscalar and the charged scalars are degenerate in mass at tree-level and are ${\mathcal{O}} (M_{KK})$.  There are no tree-level flavor changing neutral currents (FCNC) for the down-type quarks and the standard model results hold there.  In contrast, FCNC decays of the t-quark larger than in the SM are expected.
\end{abstract}

\maketitle

\section{Introduction}

With the successful start up of the Large Hadron Collider (LHC) we can look forward to the discovery of the electroweak symmetry breaking mechanism.  While the Standard Model (SM) with only one Higgs doublet is highly successful phenomenologically, it is theoretically unsatisfactory due to the Higgs boson mass being quadratically sensitive to unknown physics at high energies.  This is the hierarchy problem and is the main motivation for many models that extend the SM.  Common to many models is an extended Higgs sector.  For example, the supersymmetric extension of the SM which addresses the hierarchy problem requires two Higgs doublets, distinguished by their opposite $U(1)_Y$ charges.

Another proposed solution to the hierarchy problem is the introduction of an extra dimension with a warp factor~\cite{RS}.  These constructions, referred to as Randall-Sundrum (RS) models, are also interesting flavor models when fermions become bulk fields~\cite{flavour}.  The observed fermion mass hierarchy is given by the same warp factor which determines the electroweak-Planck hierarchy: the mass matrices are determined by fermion locations in the bulk, without fine tuning the Yukawa couplings. Generically the Klauza-Klein (KK)  excitations have masses in the TeV range and above in order to satisfy the many electroweak precision measurements.  The lower limit of $\sim$~2 to 3~TeV for KK gauge bosons make these models testable at the LHC.

Many recent studies within this RS framework of flavor reveal that the custodial symmetry of the SM Higgs potential that protects the $T$ parameter from large corrections is not automatic in 5D warped models~\cite{nocustody}.  An elegant solution is to add an $SU(2)_R$ custodial gauge symmetry which maintains tree level protection of $T$~\cite{ADMS}.  With an additional discrete left-right symmetry $P_{LR}$ the $Zb_Lb_L$ coupling will also be protected from large corrections~\cite{ACDP}.  Moreover, sizable deviations of the $Ztt$ and $Wtb$ couplings from the SM are expected and will be something to search for at the LHC.

It is a generic feature of RS flavor models that the t-quark, being much heavier than the other fermions, is located near the TeV (or IR) brane.  The KK gauge bosons also have profiles peaking close to the IR brane.  In particular, the KK gluons, $G_{KK}$, are expected to have large wavefunction overlaps with t-quarks of both handedness.  Thus in terms of the 4D effective theory the interactions $G_{KK}t\bar{t}$ are of strong coupling strength.  The exact magnitude depends on the  parameters $c^3_L,c^3_R$  that characterize the location of $t_L$ and $t_R$ in warped space.  At low energies, i.e.\ below $M_{KK}$ (the mass of the first $G_{KK}$) we might expect the t-quarks to form Nambu-Jona-Lasinio condensates \cite{Nambu}.  Specifically we examine the condensing of the SM third family doublet with $t_R$ via $G_{KK}$ exchanges to form a composite Higgs doublet.  We do not expect the lighter quarks to similarly condense since they are located further away from the IR brane, and hence the effective couplings with the $G_{KK}$ are much smaller.  This observation is not new and has been exploited to generate the Planck scale in~\cite{BCP}. A similar condensate of a fourth generation has also been studied in \cite{BR}.  Here we study the physics  of this composite doublet using the a minimal custodial RS framework described in Ref.~\cite{ADMS}. For a review of RS models see \cite{reviews}.
Ref.\cite{BH} gives a thorough review of  top quark  condensates physics.

In this paper we also assume that there is the usual SM Higgs
doublet, $H$, which is localized on the IR brane.\footnote{A
similar idea has previously been studied in Ref.~\cite{CLJK,
Aranda:2007tg}; however, both models lacked an explicit UV
completion of the 2HDM, and further in \cite{CLJK} the top quark
only coupled to the composite Higgs.}  It has the usual Yukawa
couplings to the SM fermions.  Below $M_{KK}$ the effective theory
has both an `elementary' and a composite Higgs, $\Phi$.
Electroweak symmetry breaking will involve both these scalars.
Since $\Phi$ only couples to $t_R$ there are no flavor changing
neutral currents (FCNC) in the down type quark sector.  We do
expect FCNC involving the t quarks to be important.  The t-quark
mass receives contributions from both the elementary and the
composite Higgs, so one can move the localization of $Q_3$ away
from the IR brane and still get a good agreement with the
experimental value.  This ameliorates the $Zb_Lb_L$ problem and we
find that the extra $P_{LR}$ is not always required.

Another interesting feature of the model is the simplicity of the effective scalar potential $V(\Phi,H)$.  Whereas a generic 2HDM potential contains fourteen real parameters, our model contains only eight (four of which only appear in one of two specific combinations).  Further, all of these parameters occur already in the RS framework of Ref.~\cite{ADMS}; they include the AdS curvature $k$, the 5D size $r_c$ (or equivalently the mass $M_{KK}$), the two $c$ parameters, and the 5D Yukawa couplings of the elementary scalar.  The relative simplicity of the scalar sector makes our model quite predictive; for example, the pseudoscalar and the charged Higgs are degenerate in mass at tree level.  We also find that this extended scalar sector automatically respects the $U(1)_{em}$ and CP symmetries.

In section~\ref{sec:condensate} we describe the condensate formation using $G_{KK}$ and construct $V(\Phi,H)$ which is valid  below $M_{KK}$.  In Sec.~\ref{sec:EWSB}, we study the pattern of electroweak symmetry breaking and the spectrum of scalar states.  Numerical results are given in Sec.~\ref{sec:TQmass}, and a discussion of flavor-changing neutral currents in Sec.~\ref{sec:FCNC}.   Finally we give our conclusions in Sec.~\ref{sec:Conc}.

\section{Condensate from Warp Space}\label{sec:condensate}

The model we study is the minimal custodial RS model (MCRS) given in~\cite{ADMS}.  Here we give a brief description of the model following the notations of~\cite{CNW}.  The bulk gauge symmetry is $SU(3)_c\times SU(2)_L \times SU(2)_R \times U(1)_X$.  The background geometry is a slice of $AdS_5$ space specified by the metric
\begin{equation}\label{Eq:metric}
\ud s^2 = G_{AB} \, \ud x^A \ud x^B = e^{-2\sigma(\phi)} \, \eta_{\mu\nu} \, \ud x^{\mu}\ud x^{\nu}-r_c^2 \, \ud\phi^2 \,,
\end{equation}
where $\sigma(\phi) = k r_c |\phi|$, $\eta_{\mu\nu} = \diag (1,-1,-1,-1)$, $k$ is the $AdS_5$ curvature, and $-\pi\leq\phi\leq\pi$.  The theory is compactified on an $S_1/(\set{Z}_2 \times \set{Z}_2')$ orbifold, with $r_c$ the radius of the compactified fifth dimension, and the orbifold fixed points at $\phi=0$ and $\phi=\pi$ correspond to the UV (Planck) and IR (TeV) branes respectively.  To solve the hierarchy problem, $k\pi r_c$ is set to $\approx 37$.  The warped down scale is defined to be $\tilde{k} = k \, e^{-k\pi r_c}$.  Note that $\tilde{k}$ sets the scale of the first KK gauge boson mass, $m^{(1)}_{gauge} \approx 2.45\tilde{k}$, which determines the scale of the new KK physics.

The SM quarks are embedded into $SU(2)_L \times SU(2)_R \times U(1)_X$ via the five-dimensional (5D) bulk Dirac spinors
\begin{equation}\label{Eq:qrep}
Q_i =
\begin{pmatrix}
u_{iL}\,[+,+] \\
d_{iL}\,[+,+]
\end{pmatrix} \,,\quad
U_i =
\begin{pmatrix}
u_{iR}\,[+,+] \\
\tilde{d}_{iR}\,[-,+]
\end{pmatrix} \,,\quad
D_i =
\begin{pmatrix}
\tilde{u}_{iR}\,[-,+] \\
d_{iR}\,[+,+]
\end{pmatrix} \,,\qquad
i = 1,\,2,\,3 \,,
\end{equation}
where $Q_i$ transforms as $(2,1)_{1/6}$, and $U_i$, $D_i$ transform as $(1,2)_{1/6}$.  The parity assignment $\pm$ denotes the boundary conditions applied to the spinors on the $[\mathrm {UV}, \mathrm {IR}]$ brane, with $+$ ($-$) being the Neumann (Dirichlet) boundary conditions.  Only fields with the [+,+] parity contain zero-modes that do not vanish on the brane.  These survive in the low energy spectrum of the 4D effective theory, and are identified as the SM fields.

The profiles of the zero-mode fermions are given by their flavor functions
\begin{equation}
f^0_{L,R}(\phi,c_{L,R}) = \sqrt{\frac{k \pi r_c \, (1 \mp 2c_{L,R})}{e^{k \pi r_c \, (1 \mp 2c_{L,R})}-1}} \ e^{(1/2 \mp c_{L,R})k r_c \phi} \,,
\end{equation}
where the $c$-parameter is determined by the bulk Dirac mass parameter, $m = c\,k$, and the upper (lower) sign applies to the LH (RH) label.  One of the fermion chiralities is projected out of the zero-mode; which one depends on the fermion's orbifold parity.

With the introduction of an IR brane localized Higgs field $H$ the Yukawa interactions are all localized there.  After spontaneous electroweak symmetry breaking the contribution to the mass matrices  of the SM fermions are
\begin{equation}\label{eq:RSM}
(M^{RS}_f)_{ij} = \frac{v_H}{\sqrt{2} k \pi r_c} \, y^f_{5,ij} \,
f^0_{L}(\pi,c^{L}_{f_i}) f^0_{R}(\pi,c^{R}_{f_j})
\equiv \frac{v_H}{\sqrt{2} k \pi r_c} \, y^f_{5,ij} \, F_L(c^{L}_{f_i}) F_R(c^{R}_{f_j})
\,, \qquad f = u,\,d \, ,
\end{equation}
where the label $f$ denotes up-type or down-type quark species.  $y^f_{5ij}$ denotes the 5D Yukawa couplings, $ij$ are family symbols, and $v_H$ is the VeV of $H$.

In the gauge (weak) eigenbasis, the coupling of the $n$th level KK gluon, $G^{(n)}$, to zero-mode fermions is given by
\begin{equation}
G^{A(n)}_\mu \left[ \sum_i (g^n_f)^L_{ii} \, \bar{f}_{iL} T^A \gamma^\mu f_{iL} + (L \ra R) \right] , \qquad f = u,\,d \,,
\end{equation}
where $i$ is a generation index, $T^A$ are the generators of $SU(3)$, and $(g^n_f)_{ii} = \diag (g^n_{f_1},g^n_{f_2},g^n_{f_3})$ is the weak eigenbasis coupling matrix
\begin{equation}
\label{eq:geff}
g^n_{f_i} = \frac{g_s}{\pi} \int^\pi_{0} \!\ud\phi\, |f^0(\phi,c_{f_i})|^2\chi_n(\phi) \,, \qquad g_s = \frac{g_{5s}}{\sqrt{r_c\pi}} \,.
\end{equation}
Here, $g_{5s}$ is the bulk 5D $SU(3)$ gauge coupling, $g_s$ that in the SM, and $\chi_n$ the profile of the $n$th KK gluon.  We will exclusively consider the coupling of the $G^1_{KK}$ to the t-quarks and thus write
\begin{subequations}\label{eq:gtt}
\begin{align}
g_L & = \frac{g_s}{\pi} \int^\pi_{0} \!\ud\phi\, |f^0(\phi,c_{f_{3L}})|^2 \chi_1(\phi) \\
g_R & = \frac{g_s}{\pi} \int^\pi_{0} \!\ud\phi\, |f^0(\phi,c_{f_{3R}})|^2 \chi_1(\phi) .
\end{align}
\end{subequations}
For processes with small external momenta, tree-level exchange of $G^1_{KK}$ leads to 4-Fermi interactions between zero mode fermions given by
\begin{equation}
\label{eq:cond}
-\frac{g_i g_j}{M_{KK}^2}\left(\overline{Q_{iL}}T^A\gamma^\mu Q_{iL}\right)\left(\overline{f_{jR}}T^A\gamma_\mu f_{jR}\right)
=\frac{g_i g_j}{M_{KK}^2}\left(\overline{Q_{iL}} f_{jR}\right)\left(\overline{f_{jR}}Q_{iL}\right) + O(1/N_c)
\end{equation}
where $N_c=3$ for $SU(3)_c$. Taking  $Q_i=Q_3$ and $f_j=t_R$, this generates the t-quark condensate considered in~\cite{Nambu}.

The exchange of spin-1 KK-gluons leads to an attractive force in Eq.~(\ref{eq:cond}).  A condensate can only form if the effective couplings' $g_i$'s are sufficiently strong.  In the RS model this requires the overlap of the KK gluons and the fermion zero modes be large and thus enhance the strong coupling $g_s$.  The wavefunction of the lightest KK gluon $G^1_{KK}$ peaks near the IR brane; so the only fermions that can condense are those which are similarly localized.  Eq.~\eqref{eq:RSM} shows that light fermions are oriented towards the UV brane, and so no enhancement of $g_s$ is expected.  The large top mass requires the fields $Q_3$ and $t_R$ to be IR localized; as we shall see later, this remains true in the presence of the composite Higgs, since a substantial part of the t-quark mass still comes from $H$.  This justifies our restriction of Eq.~\eqref{eq:cond} to only $Q_3$ and $t_R$.

We have focused on the role of $G^1_{KK}$ in condensate formation.  The higher KK-gluon modes contribute only a small correction, due both to a smaller overlap integral from Eq.~(\ref{eq:geff}) and their higher masses.  Numerical calculations show that the contribution of these states to the coefficient in Eq.~\eqref{eq:cond} is at most $\sim 5\%$ for parameters of interest (and usually much less).  Similarly, the KK fermions give small corrections to the above picture when they are integrated out.

Below the scale of $M_{KK}$ the condensate in Eq.~({\ref{eq:cond}) can be viewed as a composite Higgs doublet denoted by $\Phi$~\cite{Nambu}.  It has the same $SU(2)_L\times U(1)_Y$ quantum numbers as the SM Higgs.  This is in addition to the elementary scalar field $H$ which is in the MCRS model.  At the scale $M_{KK}$, $\Phi$ is a static auxiliary field.  Following~\cite{BHL} we can write the Lagrangian at scale $M_{KK}$
as
\begin{equation}
\label{eq:LMkk}
{\mathcal{L}}= |D_\mu H|^2 - m_0^2 H^\dagger H - \frac{1}{2} \lambda_0 (H^\dagger H)^2 +
\lambda_t \overline{Q_{L}} t_R \widetilde{H} + g_t \overline{Q_L}t_R \widetilde{\Phi}- M_{KK}^2 \Phi^\dagger \Phi + h.c.
\end{equation}
where $\widetilde{H}= i\sigma_2 H^*$, $\widetilde{\Phi}= i\sigma_2 \Phi^*$, $Q_L=(t,b)_L$, and $m_0^2$, $\lambda_0$ are the parameters in the brane Higgs scalar potential.  We have omitted the Yukawa interactions of the brane Higgs with light fermions as they play no role here.  $D_\mu$ denotes the gauge covariant derivative.  Integrating out $\Phi$ reproduces the 4-Fermi interaction of Eq.~(\ref{eq:cond}).  For the Yukawa couplings, $g_t \equiv \sqrt{g_Lg_R}$ is given in Eq.~(\ref{eq:gtt}), and $\lambda_t$ is determined by the wavefunction overlap of $Q_3$, $t_R$, and $H$ at the IR brane.  Specifically, using the notation of Eq.~\eqref{eq:RSM} and defining $y_{5D} \equiv y^u_{5,33}$ we have
\begin{equation}
\lambda_t = \frac{y_{5D}}{k\pi r_c} \, F_L (c_{3L}) \, F_R (c_{tR}) = y_{5D} \sqrt{\frac{(1-2c_{3L})(1+2c_{tR})}{(1-e^{k\pi r_c(2c_{3L}-1)})(1-e^{-k\pi r_c(2c_{tR}+1)})}} .
\label{eq:lambdat}\end{equation}

\begin{figure}
\centering
\includegraphics[width=3.0in]{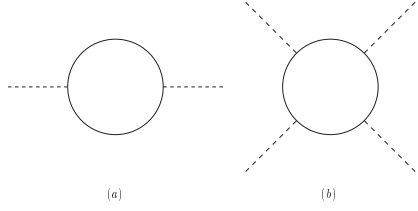}
\caption{Fermion bubble contribution to scalar (a) 2-point functions and
(b) 4-point functions. The dashed lines can be either $\Phi$ or $H$ fields.\label{fig:bub}}
\end{figure}

At scales $\mu < M_{KK}$, quantum fluctuations generate a kinetic term for $\Phi$ as well as kinetic and mass term mixings between $\phi$ and $H$. We will calculate these corrections in the fermion bubble approximation, depicted in Fig.~\ref{fig:bub}.  The effective Lagrangian takes the form
\begin{equation}
\label{eq:Lmu}
\begin{split}
{\mathcal{L}} & = \left[ 1 + \lambda_t^2 \epsilon \right] |D_\mu H|^2 + \lambda_t g_t \epsilon \left[ (D_\mu H)^\dagger D^\mu \Phi + h.c. \right] + g_t^2 \epsilon |D_\mu \Phi|^2 \\
& \quad - \left[ m_0^2 - \lambda_t^2 \Delta^2 \right] H^\dagger H + \lambda_t g_t \Delta^2 \left[ H^\dagger \Phi + \Phi^\dagger H\right] - \left[ M_{KK}^2 - g_t^2 \Delta^2 \right] \Phi^\dagger \Phi \\
& \quad - \left[ \frac{1}{2} \lambda_0 + \lambda_t^4 \epsilon \right](H^\dagger H)^2
- 2 \lambda_t^2 g_t^2 \epsilon \left[ \frac{1}{2} (H^\dagger \Phi +\Phi^\dagger H)^2 + H^\dagger H \Phi^\dagger \Phi \right] \\
& \quad - 2 \lambda_t^3 g_t \epsilon \, H^\dagger H (H^\dagger \Phi +\Phi^\dagger H) - 2 \lambda_t g_t^3 \epsilon \, \Phi^\dagger \Phi (H^\dagger \Phi +\Phi^\dagger H) - g_t^4 \epsilon \, (\Phi^\dagger \Phi)^2 \\
&\quad + \lambda_t \overline{Q_L} t_R \widetilde{H} + g_t \overline{Q_L} t_R \widetilde{\Phi} + h.c.
\end{split}
\end{equation}
Here,
\begin{equation}
\begin{split}
\epsilon & = \frac{N_c}{16\pi^2} \ln \left(\frac{M_{KK}^2}{\mu^2}\right) ; \\
\Delta^2 & = \frac{2N_c}{16\pi^2} \bigl( M_{KK}^2-\mu^2 \bigr) ,
\end{split}
\end{equation}
are calculated in the 1-loop approximation. We have also taken the cutoff to be $M_{KK}$, above which the 4-Fermi condensate approximation is no longer valid.

The kinetic mixing term can be diagonalized in the customary fashion.  The transformations
\begin{equation}
\label{eq:kmixing}
\begin{split}
H & =\hh\\
\Phi & = - \frac{\lambda_t}{g_t} \hh + \frac{1}{g_t\sqrt \epsilon} \hphi
\end{split}
\end{equation}
will cast the kinetic terms into canonical diagonalized form. The resulting Lagrangian of the scalars is delightfully simple:
\begin{equation*}
{\mathcal{L}} \supset |D_\mu \hh|^2 + |D_\mu \hphi|^2 - V(\hh,\hphi)
\end{equation*}
where the scalar potential is given by
\begin{equation}
\label{eq:pot}
\begin{split}
V(\hh,\hphi) & = \left( m_0^2 +\frac{\lambda^2_t}{g_t^2}M^2_{KK}\right) \hh^\dagger \hh - \frac{\lambda_t}{g^2_t\sqrt \epsilon} M_{KK}^2 \left( \hh^\dagger \hphi +\hphi^\dagger\hh \right) \\
& \quad + \left(\frac{M_{KK}^2}{g_t^2\epsilon} - \frac{\Delta^2}{\epsilon}\right) \hphi^\dagger \hphi + \frac{1}{2}\lambda_0 (\hh^\dagger\hh)^2 + \frac{1}{\epsilon} (\hphi^\dagger \hphi)^2
\end{split}
\end{equation}
In this basis the only interaction between the two scalar doublets is the mass mixing.  This mixing is related to the $c$ parameters via $g_t$ and $\lambda_t$.  We reiterate that this potential is valid in the range $\mu < M_{KK}$.

\section{ElectroWeak Symmetry Breaking of 2HDM}\label{sec:EWSB}

The particular form of our Higgs potential makes study of the scalar VeVs and physical states straightforward.  It is notable that in this model, alignment of the Higgs VeVs is automatic from the structure of the potential; while CP is conserved since all parameters are real.  We denote the vacuum expectation values of $\hh$ and $\hphi$ by $v_H$ and $v_\phi$ respectively, and also define $\tan \beta =\frac{v_H}{v_\phi}$.  Minimizing $V(\hh,\hphi)$ yields two coupled cubic equations
\begin{equation}
\left(m_0^2 +\frac{\lambda_t^2}{g_t^2}M_{KK}^2\right) v_H - \frac{\lambda_t}{g_t^2 \sqrt \epsilon}M_{KK}^2 v_\phi +
\frac{ \lambda_0}{2}| v_H|^2v_H=0,
\label{eq:min1}\end{equation}
and
\begin{equation}
\left( \frac{M_{KK}^2}{g_t^2\epsilon} - \frac{\Delta^2}{\epsilon}\right) v_\phi - \frac{\lambda_t}{g^2_t\sqrt \epsilon}M_{KK}^2 v_H + \frac{2}{\epsilon}|v_\phi|^2v_\phi=0 .
\label{eq:min2}\end{equation}
We require $v_H^2 +v_\phi^2 =v^2$ where $v=246$ GeV.

The physical scalar spectrum consists of a pair of charged Higgs $H^\pm$, a pseudoscalar $A$, and two neutral scalars $h_{1,2}$.  The charged and pseudoscalar sectors have the same mass matrix
\begin{equation}
\label{eq:Mpm}
M^2_{\pm}=
\begin{pmatrix}
a + \frac{\lambda_0}{2} v^2_H & c \\
c & b  +\frac{1}{\epsilon} v^2_\phi
\end{pmatrix},
\end{equation}
where
\begin{align}
a & = m_0^2 + \frac{\lambda_t^2}{g_t^2} M^2_{KK} ; \\
b & = \frac{1}{\epsilon} \biggl( \frac{M^2_{KK}}{g_t^2}-\Delta^2 \biggr) ; \\
c & = -\frac{\lambda_t}{g^2_t\sqrt{\epsilon}} M^2_{KK} .
\end{align}
Since $c^2 = (a+\frac{\lambda_0}{2}v_H^2)(b+\frac{1}{\epsilon}v^2_\phi)$, this matrix has a null eigenvalue, corresponding to the eaten Goldstone modes.  The non-zero eigenvalue gives the mass of the physical states, and hence at tree level we have
\begin{equation}
M_A^2 = M_H^2 = \frac {2\lambda_t}{g_t^2\sqrt {\epsilon}\sin {2\beta}}M_{KK}^2
\label{eq:chhiggs}\end{equation}
We shall see in section~\ref{sec:TQmass} that these masses are expected to be $\lesssim M^2_{KK}$.

The degeneracy of the charged and pseudoscalar Higgses can be understood from a symmetry stand point.  The 2HDM has the maximal symmetry of $SO(8)$.  Without the mass mixing term the $\hphi$ and $\hh$ will have the individual symmetries $SU(2)_{\hphi L}\times SU(2)_{\hphi R}$ and $SU(2)_{\hh L}\times SU(2)_{\hh R}$, and their cross product is a subgroup of $SO(8)$.  Spontaneous symmetry breaking yields
\begin{align}
SU(2)_{\hphi L}\times SU(2)_{\hphi R} & \stackrel{v_{\phi}}\longrightarrow SU(2)_{D\hphi} \,; \notag \\
SU(2)_{\hh L}\times SU(2)_{\hh R}&\stackrel{v_{H}}\longrightarrow SU(2)_{D\hh} \,,
\end{align}
and the $SU(2)_D$ are analogous to the custodial $SU(2)$ of the SM scalar potential after symmetry breaking.  These symmetries are then softly broken by the mixing term in $V(\hphi,\hh)$, which indicates that there is a remaining custodial symmetry $SU(2)_V \subset SU(2)_{D\hphi}\times SU(2)_{D\hh}$.  The three Goldstones form a triplet under this $SU(2)_V$.  The two charged Higgs and the pseudoscalar form another triplet.  The remaining two scalars are singlets and complete the number of fields.  Hence, at tree level we have $M_A= M_H$.

The mass squared matrix of the two scalars is given by
\begin{equation}
\label{eq:M20}
M^2_0 =
\begin{pmatrix}
a + \frac{3}{2} \lambda_0 v_H^2 & c\\
c & b + \frac{3}{\epsilon} v^2_\phi
\end{pmatrix}
\end{equation}
Taking the trace and the determinant of this matrix gives $\Tr M^2_0 = M_H^2 + k_1 v^2$ and $\det M^2_0 = k_2 M^2_H v^2$, where $k_{1,2} $ are ratios of ${\mathcal{O}} (1)$ parameters.  This implies that one of the scalars has a mass $\sim M_H \sim {\mathcal{O}} (\mathrm {TeV})$, while the other has mass of order of electroweak scale.  Heavy scalars are expected in this model.  This can be seen from the scalar potential given in Eq.~(\ref{eq:pot}), since the quartic term $(\hphi^\dagger \hphi)^2 $ has a large coefficient. We close this section by giving the mixing angle $\alpha$ between the two scalars:
\begin{equation}
\label{eq:hmixing}
\tan {2\alpha} =-\frac{2c}{a+\frac{3}{2}\lambda_0 v_H^2 -b-\frac{3}{\epsilon}v^2_{\phi}},
\end{equation}
and
\begin{equation}
\left(\begin{array}{c}  H_0 \\h_0  \end{array}\right)=
\left(\begin{array}{cc} \cos\alpha & -\sin\alpha \\ \sin\alpha & \cos\alpha   \end{array}\right)
\left(\begin{array}{c} \hat{H}_R \\ \hat{\Phi}_R   \end{array}\right) .
\label{eq:scamixing}\end{equation}
The subscript $R$ stands for the real part expansion around the VeVs, and $h_0$ ($H_0$) is the physical lighter (heavier) neutral scalar.  We will show in section~\ref{sec:TQmass} that the couplings of $h_0$ to gauge bosons are almost identical to those of the SM Higgs; we are in the `decoupling' limit of 2HDMs~\cite{2HDM:dec}.

\section{t-quark mass and numerical analysis}\label{sec:TQmass}

We start with a discussion of the top quark mass generation.  Below the cut-off scale of $M_{KK}$ the 2HDM is valid.  After making the field rotation of Eq.~\eqref{eq:kmixing}, the Yukawa sector for the 3rd generation becomes
\begin{equation}
\mathcal{L}_Y = \lambda_t \overline{Q_L} t_R \widetilde{H} + g_t \overline{Q_L} t_R \widetilde{\Phi} + h.c. \to \frac{1}{\sqrt{\epsilon}} \overline{Q_L} t_R \widetilde{\hphi} + h.c.
\label{eq:3genyuk}\end{equation}
Eq.~\eqref{eq:3genyuk} shows that the t-quark gets its mass from coupling to $\hphi$, which after symmetry breaking gives\footnote{We are neglecting the contributions from Yukawas involving $H$ that mix generations, since these terms will be small.}
\begin{equation}
\label{eq:mt}
m_t = \frac{v\cos \beta}{\sqrt {2\epsilon}}.
\end{equation}
We have already assumed that $v$ has the SM value; Eq.~\eqref{eq:mt} then tells us that the top quark mass determines the value of $\beta$.  This determination of \emph{both} Higgs field VeVs is unusual in 2HDMs.

To make use of this result, we need to specify the renormalization scale $\mu$.  It is reasonable to choose $\mu=v$.  Using the value of $m_t$ at the electroweak scale fixes $\cos \beta \sim \sqrt{\epsilon}$.  Eq.~(\ref{eq:min2}) then becomes a constraint on $\lambda_t$, $g_t$, with the allowed regions shown in Fig.~\ref{fig:ygsol}.  Recall that $g_t=\sqrt{g_Lg_R}$ is related to the parameters $c_L^3$, $c_R^3$ via Eq.~\eqref{eq:gtt}, and $\lambda_t$ is related to them via Eq.~\eqref{eq:lambdat}; so for a given $y_{5D}$, this becomes a restriction on the 5D mass parameters of the third generation.

\begin{figure}%[h]
\centering
\includegraphics[height=0.3\textheight]{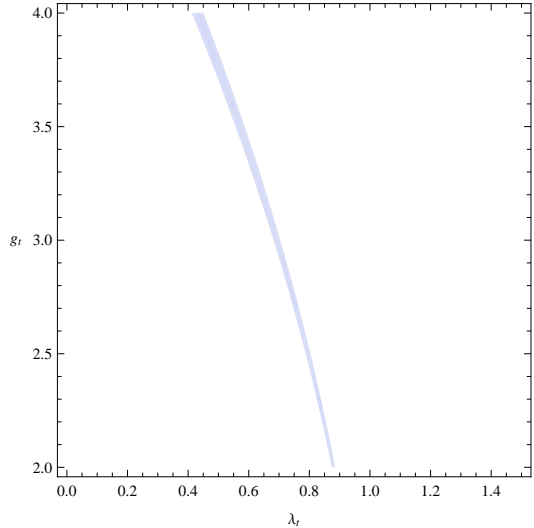}
\caption{Allowed region in the $\{ \lambda_t$, $g_t\}$ parameter space that satisfies Eqs.~\eqref{eq:mt} and \eqref{eq:min2}.  Note that for solutions in the region shown, we must take $\cos \beta$ positive. The spread corresponds to taking $M_{KK}$ to lie between 1.5 to 4 TeV and experimentally allowed region of $m_t$ from 169.7 to 172.9 GeV.\label{fig:ygsol}}
\end{figure}

Taking into account the above considerations,  Eq.~(\ref{eq:min1}) now constrains the $m_0^2$ and $\lambda_0$ parameters of the brane Higgs; see Fig.~\ref{fig:lmsol} for the results.  We see that while $m_0$ is of the order of the weak scale the quartic coupling of the brane Higgs field is of ${\mathcal{O}}(1)$ and not smaller; but the coupling remains well within the perturbative regime.  We also see that, after imposing the various constraints, there remain three free parameters in our 2HDM, which we take to be $g_t$, $\lambda_0$ and $M_{KK}$.

\begin{figure}%[h]
\centering
\includegraphics[height=0.3\textheight]{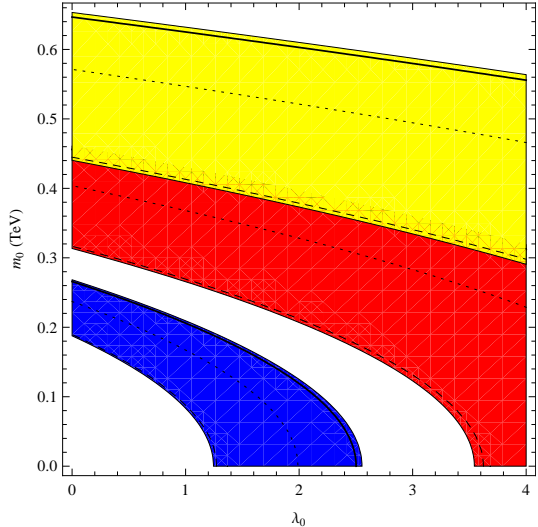}
\caption{Allowed region in the $\{ \lambda_0$, $m_0\}$ parameter space that satisfies Eq.~\eqref{eq:min1}. The blue, red and yellow regions correspond to $M_{KK}= 1.5, 2.5\; {\mathrm{and}}\; 3.5$ TeV respectively.  The different lines (solid, dotted, dash)  correspond to different values of $g_t= 2,3,4$.  We have taken $\sin \beta$ positive, which implies $m_0^2$ positive, and the central experimental value for $m_t$.\label{fig:lmsol}}
\end{figure}

Numerically solving for the spectrum of the scalars, we predict the mass of the lighter Higgs boson to be in the range of 100--400~GeV; while the heavier scalar, pseudoscalar and charged Higgs are all near $M_{KK}\sim$~TeV.  We plot the lightest scalar mass in Fig.~\ref{fig:higgs} and the others in Fig.~\ref{fig:scalars}.

For $h_0$, the mass is determined almost entirely by $\lambda_0$, with essentially no dependence on $g_t$.  As Fig.~\ref{fig:higgs} demonstrates, varying $M_{KK}$ also has very little effect on the mass of this scalar.  We further note that this Higgs boson has essentially the properties of the SM Higgs; the mixing angles $\alpha$ and $\beta$ are almost equal, with the choice of $2\alpha$ in the second quadrant (see Eq.~(\ref{eq:hmixing})).  All these properties can be qualitatively understood from the mass squared matrices of the neutral and charged scalars, Eq.~(\ref{eq:Mpm}) and Eq.~(\ref{eq:M20}).  We can write
\begin{equation}
\label{M2dif}
M_0^2 = M^2_{\pm} + \begin{pmatrix} \lambda_0v^2 \sin \beta &0 \\ 0 & 4 m_t^2 \end{pmatrix}.
\end{equation}
The charged Higgs masses are ${\mathcal{O}}(M_{KK})$ and the matrix $M^2_\pm$ is diagonalized by a rotation matrix with angle $\beta$.  The second term of Eq.~(\ref{M2dif}) can then be treated as a perturbation, being of ${\mathcal{O}}(.01)$.  This implies $\beta\simeq \alpha$, that one of the two scalars should be nearly degenerate with the charged/pseudoscalar Higgses, and that the other should be at the electroweak scale.  Using Eq.~(\ref{M2dif}) we find an approximate expression for the mass of $h_0$:
\begin{equation}
\label{eq:h0mass}
M^2_{h_0}\simeq \lambda_0 v^2 \sin ^4 \beta + 2\epsilon \, m_t^2.
\end{equation}

The near equality of $\alpha$ and $\beta$ implies that the lighter scalar is very SM like.  For example, its coupling to the Z boson has a factor of $\cos (\beta-\alpha)$ relative to the SM coupling; so it has essentially SM strength.  The LEP bound on the Higgs mass of $>114$~GeV~\cite{LEP}, as well as the Tevatron bound $m \notin (163,166)$~GeV~\cite{Tev}, apply and are shown in Fig.~\ref{fig:higgs}.  A first glance at Eq.~\eqref{eq:3genyuk} would suggest that the light scalar would have no coupling to t-quarks.  However, we will show in section~\ref{sec:FCNC} that the couplings of $h_0$ to all fermions are proportional to $m_f \sin\alpha /\sin\beta \simeq m_f$.  We conclude that the lighter scalar is indistinguishable from the SM Higgs.

\begin{figure}%[h]
\centering
\includegraphics[height=0.3\textheight]{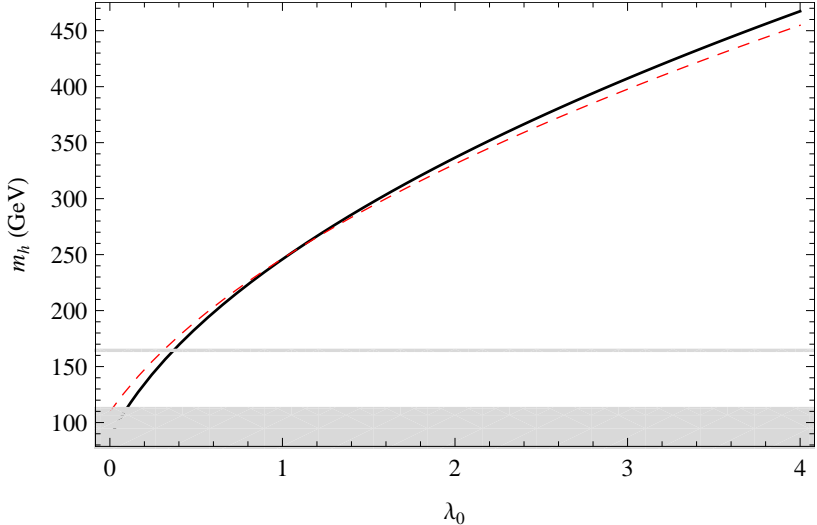}
\caption{Value of the mass of the lighter Higgs boson, as a function of $\lambda_0$. The black solid line is for $M_{KK}= 1.5$ TeV and the red dashed line is for $M_{KK}=4$ TeV. The shaded regions are the LEP and Tevatron exclusions for the Higgs mass. \label{fig:higgs}}
\end{figure}

For the heavier scalars, the masses are mostly determined by $g_t$, exhibiting little dependence on $\lambda_0$.
This is congruent with our expectations from Eq.~\eqref{eq:chhiggs} and Eq.~\eqref{M2dif}.  Note that as shown in Fig.~\ref{fig:scalars}, all these states are nearly degenerate with the scalar slightly heavier then the charged and the pseudoscalar states. The splittings are of ${\mathcal{O}}(10)$ GeV.  We also note that the heavier scalar coupling to the Z-boson is negligible, since it is proportional to $\sin (\beta-\alpha)$.  This state can be produced at the LHC if kinematically allowed, but will not be produced in a $e^+e^-$ collider in the $ZH$ channel.

\begin{figure}%[h]
\centering
\includegraphics[height=0.3\textheight]{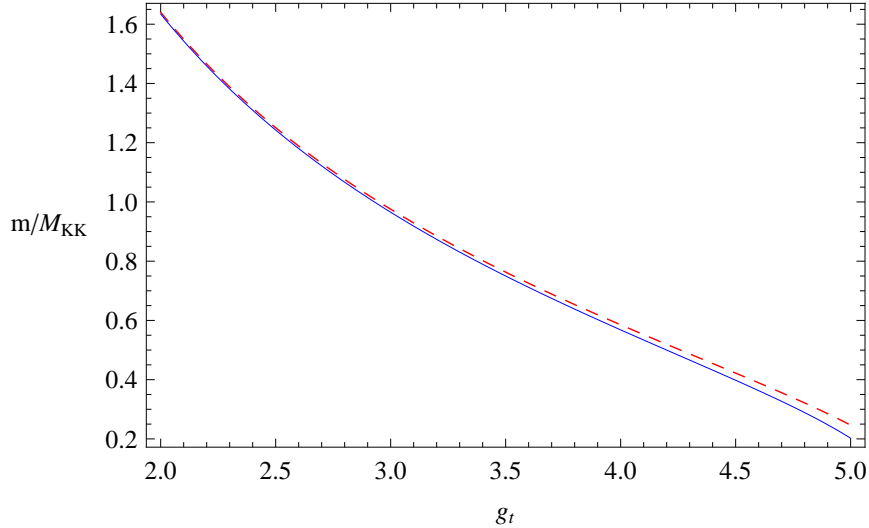}
\caption{The masses of the charged, pseudoscalar and heavier scalar Higgses, as a function of $g_t$ and relative to $M_{KK}$.  The charged and pseudoscalar states are degenerate at tree level and are shown by the blue, solid line; the heavier scalar state is the red, dashed line.\label{fig:scalars}}
\end{figure}

An alternative way of obtaining $m_t$ is via the gap equation with a brane Higgs contribution.
\begin{figure}%[h]
\centering
\includegraphics[width=4.5in]{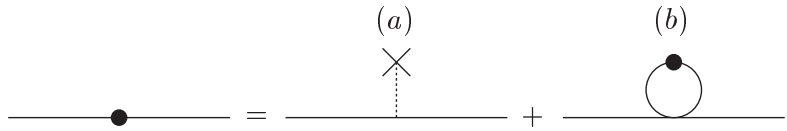}
\caption{The gap equation for $m_t$.  (a) Contribution from the brane Higgs (dash line) and  the cross denotes the VeV.  (b) The fermion bubble contribution.\label{fig:gappic}}
\end{figure}
Instead of Eq.(\ref{eq:mt}) we have
\begin{equation}
\begin{split}
m_t&= \frac{\lambda_t}{\sqrt 2}v\sin \beta -i \frac{N_cg_Lg_R}{2M_{KK}^2}\int \frac{d^4\ell}{(2\pi)^4}\frac{{\mathrm {Tr}}(\ell\!\!\! / +m)}{\ell^2- m_t^2}\\
&=\frac{\lambda_t}{\sqrt 2}v\sin \beta +  \frac{N_cg_Lg_Rm_t}{8\pi^2}\left[ 1+\frac{m_t^2}{M_{KK}^2}\ln \frac{m_t^2}{M_{KK}^2} \right]\\
\end{split}
\end{equation}
The graphical depiction of this is given in Fig.~\ref{fig:gappic}.  This approach includes non-perturbative effects via the fermion loop.  The range of masses predicted for $h_0$  is between 110 to 470 GeV which is very close to that from the 2HDM (see Fig.~\ref{fig:higgs}). We find that the approaches agree with each other to within twenty percent in the observables we calculated.  This is consistent with the approximation of dropping $1/N_c$ terms.  Fig.~\ref{fig:gap_tanB} is  an example of the qualitative agreement and quantitative difference of the two calculations.
\begin{figure}[h]
\centering
\includegraphics[width=2.5in]{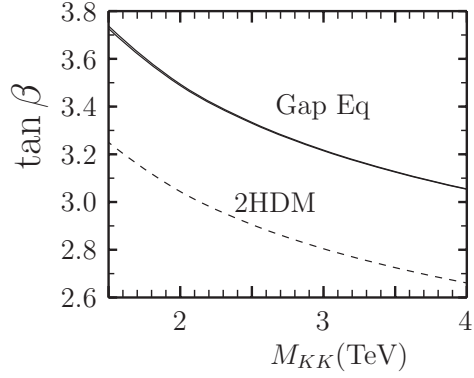}
\caption{The solutions for $\tan\beta$ v.s. $M_{KK}$ in the 2HDM approach and from the gap equation.
\label{fig:gap_tanB}}
\end{figure}

Although it is difficult to quantify at what coupling strength fermion condensates will form, we can draw from our experience with low energy QCD.  We expect condensates to form if the effective coupling $g_t\gtrsim g_s(2~{\text{GeV}}) \sim$~2.1.   Our solutions shown in Fig.~\ref{fig:ygsol} are consistent with that.  In Fig.~\ref{fig:gclcr} we show how the effective condensate coupling $g_t$ is related to the $c$-parameters.
\begin{figure}%[h]
\centering
\includegraphics[width=3.5in]{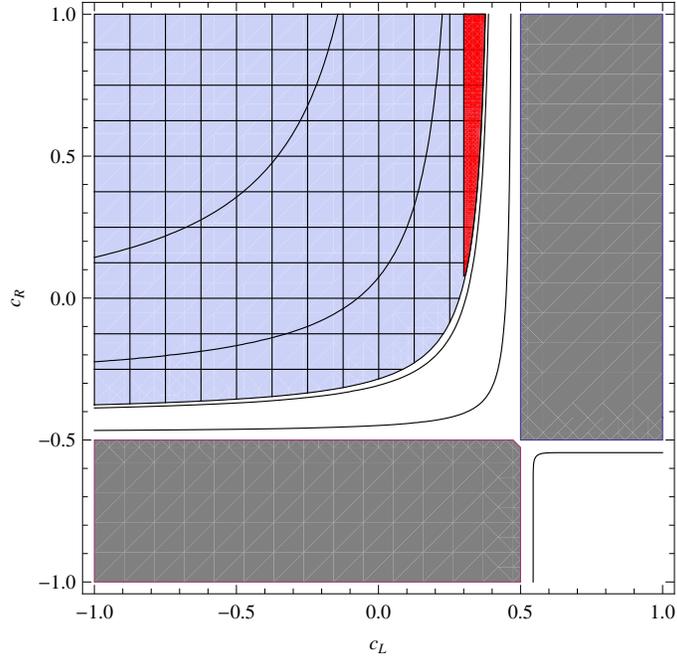}
\caption{Contours of $g_t$ in $c_L^3, c^3_R$ plane for $M_{KK}= 1.5$ TeV. The solid lines are $g_t$ contours for $g_t = \{ 4,3,2,1,0.125\}$ from top-left to bottom right.  The dark regions are where $g_L$ and $g_R$ have opposite signs, and hence are inconsistent with the condensate scenario.  The small red region gives a good fit to $Z\ra b_L\bar{b}_L$, without the additional $P_{LR}$ symmetry.\label{fig:gclcr}}
\end{figure}
In order to satisfy both the EWPT and the constraints from the correction to the SM $\delta g^L_b< 0.01$ we require $c_L^3 > 0.3$ \cite{ADMS,CNW} and this is marked in Fig.~\ref{fig:gclcr}. In Fig.~\ref{fig:gapcLcR} we show the allowed region and the contour of 5D Yukawa couplings that gives the t-quark mass.
\begin{figure}%[h]
\centering
\includegraphics[width=3.0in]{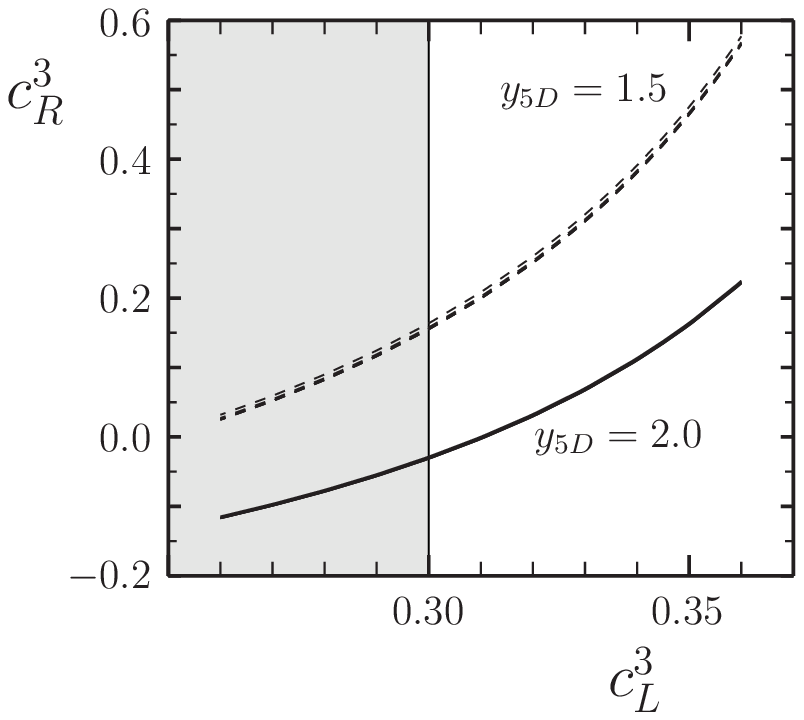}
\caption{The solution for bulk mass parameters $c^3_L$ and $c^3_R$ with two representative 5D Yukawa couplings.  The KK mass is varied from 1.5 TeV to 4.0 TeV.
The shaded areas are excluded by the $Z\ra b_L\bar{b}_L$.
\label{fig:gapcLcR}}
\end{figure}
Hence, we conclude that, with the aid of a t-quark condensate, the parameter space for  the minimal custodial RS model does not require a further discrete symmetry to satisfy
all the phenomenological constraints in a small region. Most of the parameter space will require a discrete symmetry in order to accommodate the $Z\ra b_L\bar{b}_L$ constraint.

\section{Flavor Changing Neutral Current Effects.}\label{sec:FCNC}

Now we consider the full Yukawa sector for the 2HDM model, including light quarks.  The Lagrangian is given by
\begin{equation}
{\mathcal {L}}_Y= \lambda^d_{ij} \overline{Q_{Li}}d_{jR} H + g_t\overline{Q_{3L}}t_R \widetilde{\Phi} + \lambda^u_{ij}\overline{Q_{iL}}u_{jR} \widetilde{H} + h.c.
\end{equation}
where ${i,j}$ are family indices and other notations are standard.  The condensate $\Phi$ only couples to the third family.  After the rotation to remove scalar kinetic mixing given in Eq.~(\ref{eq:kmixing}), we get
\begin{equation}
\label{eq:LY}
{\mathcal{L}}_Y = \lambda^d_{ij} \overline{Q_{Li}}d_{jR} \hh + \left( \lambda^u_{ij} - \lambda^u_{33} \right)\overline{Q_{iL}}u_{jR}\widetilde{\hh} + \frac{1}{\sqrt \epsilon}\overline{Q_{3L}}t_R \widetilde{\hphi} + h.c.
\end{equation}

After spontaneous symmetry breaking, the mass matrices are given by replacing $\hh\rightarrow v_H/\sqrt{2}$ and $\hphi \rightarrow v_\phi/\sqrt{2}$.  The down type mass matrix is proportional to $\lambda^d_{ij}$, so diagonalizing this also diagonalizes the Yukawa matrix.  There are therefore no tree level flavor changing neutral currents for the $d$-type quarks.

However, the up type quarks receive masses from VeVs of both doublets.  This implies that when we diagonalize the $u$-quark mass matrix, we will not simultaneously diagonalize the Yukawa matrix.  This leads to tree level FCNC, which are obviously a matter of concern.  While we defer a full study of these effects for later work, here we argue that they will not be intolerable.  First, we can write the mass matrix for up-type quarks as
\begin{equation}
\mathcal{M}^u_{ij} = -\frac{1}{\sqrt{2}}\begin{pmatrix}
\lambda^u_{11} v_H & \lambda^u_{12} v_H & \lambda^u_{13} v_H \\
\lambda^u_{21} v_H & \lambda^u_{22} v_H & \lambda^u_{23} v_H \\
\lambda^u_{31} v_H & \lambda^u_{32} v_H & \frac{1}{\sqrt{\epsilon}} v_\phi
\end{pmatrix} .
\end{equation}
This lets us rewrite the Yukawa sector \eqref{eq:LY} as
\begin{equation}
{\mathcal{L}}_Y = -\frac{\sqrt{2}\mathcal{M}^d_{ij}}{v \sin\beta} \overline{Q_{Li}}d_{jR} \hh -\frac{\sqrt{2}\mathcal{M}^u_{ij}}{v \sin\beta} \overline{Q_{iL}}u_{jR}\widetilde{\hh} + \frac{1}{\sqrt \epsilon} \overline{Q_{3L}} t_R \left( \widetilde{\hphi} - \frac{\widetilde{\hh}}{\tan\beta} \right) + h.c.
\label{eq:FCNChiggs}\end{equation}
It is manifestly clear that FCNCs arise \emph{only} from the last term.

There are three effects suppressing light quark FCNCs from this term.  The first two arise from the expansion of the Higgs doublets in terms of physical fields.  Note that the combination $\sin\beta\,\widetilde{\hphi} - \cos\beta\,\widetilde{\hh}$ has no VeV, and so contains purely the \emph{physical} charged and pseudoscalar fields.  Because of the near equality of the mixing angles $\alpha$ and $\beta$, it also is almost pure $H_0$.  This means that flavour changing currents through the neutral scalars are either suppressed by the relatively large mass $M_H^2$, if they proceed via $H_0$; or by the small parameter $\sin(\beta-\alpha)$, if they proceed via $h_0$.

The third effect suppressing light quark FCNCs comes from the flavor structure of RS.
The gauge basis we have used above is close to the 4D mass basis; the $c$-parameters of the 5D fermions naturally generate hierarchical Yukawa matrices.  This implies that $\overline{Q_{3L}}$ and $t_R$ states contain only small admixtures of the up and charm quarks, suppressing FCNCs that do not involve t-quarks.  We expect that the combination of these factors suffices to push new contributions from this model to measured observables to sufficiently small values.

Finally, we briefly comment on the first two terms in Eq.~\eqref{eq:FCNChiggs}.  These terms involve the dominant coupling of the light scalar $h_0$ to fermions, since the last term introduces a heavily suppressed coupling as already noted.  We note that the neutral sector of $\hh$ is $\sin\alpha\,h_0 + \cos\alpha\,H_0$ (see Eq.~\eqref{eq:scamixing}).  Then we see that the Yukawa matrix for $h_0$ is $-\sqrt{2}(\mathcal{M}^d_{ij}/v) \sin\alpha/\sin\beta$, as already claimed in section~\ref{sec:TQmass}.

\section{Conclusions}\label{sec:Conc}

We have derived an effective 2HDM from the Randall-Sundrum scenario by using a IR brane localized Higgs field and a second scalar doublet obtained from t-quark condensation induced by KK gluon exchange.  The first KK gluon mode is dominant, and we take its mass $M_{KK}$ to be the cutoff for the effective theory.  The scalar kinetic mixing and mass mixing are calculated in the one fermion bubble approximation.  After the diagonalization that gives canonical kinetic terms the resulting 2HDM is surprisingly simple.  The model has no tree level FCNC effects for the d-type quarks and hence does not run afoul of constraints from B meson and kaon decays.  On the other hand we expect sizable FCNC decays of the t-quark which can be searched for at the LHC, and we defer this and other flavor physics issues to a later work.  Moreover, one can show that the model does not support spontaneous T-violation.

We obtained the t-quark mass in two different ways.  The first was to use the effective 2HDM, which after SSB yields the result that $m_t$ is proportional to $\cos \beta$, where $\beta$ is the mixing angle of the two VeV's.  For the second approach we derive an inhomogeneous gap equation for $m_t$, using the fermion bubble approximation and the brane Higgs.  The two results are in qualitative agreement, and quantitatively they differ by no more than 20\%.  This is well within the $N_c^{-1}$ accuracy of our approximation.

As in any 2HDM, there are five physical spin-0 particles.  The model predicts that the pair of charged scalars and the pseudoscalar are degenerate at tree level.  They are expected to be close to $M_{KK}$, as is the heavier real scalar.  The remaining scalar  is at the electroweak scale, and has couplings to fermions and gauge bosons that are essentially the same as the SM Higgs.  The heavier scalar has a small coupling to the electroweak gauge bosons, and hence can only be produced at the LHC via gluon fusion through quark loops.

Since the second Higgs owes its origin to the formation of a t-condensate due to KK gluons, it is not surprising that the properties of the resulting 2HDM depend on the localization $c$-parameters of the t-quark.  We note that there is a small region of parameter space that satisfies the EWPT and the constraints of $Z\ra b_L\bar{b}_L$ in the
custodial $SU(2)_L\times SU(2)_R \times U(1)_X$ RS model without an additional discrete symmetry.

\begin{acknowledgments}
The research of W.F.C. is supported by the Taiwan NSC under Grant No.\ 96-2112-M-007-020-MY3.  The research of J.N.N. and A.P.S. is partially supported by the Natural Science and Engineering Council of Canada.  J.N.N. would like to thank Prof L.F. Li for the hospitality of the National Center for Theoretical Science where part of this work was done.
\end{acknowledgments}

\end{document}